\documentclass[1p,12pt]{elsarticle}

\usepackage{graphicx,amssymb,bm,mathrsfs,float,geometry,slashed,subcaption}
\usepackage{tikz}
\usetikzlibrary{shapes, arrows, positioning, calc}
\usepackage{amsmath} 
\numberwithin{equation}{section} 
\numberwithin{figure}{section} 
\numberwithin{table}{section}
\usepackage[dvipsnames]{xcolor}
\definecolor{forestgreen}{rgb}{0.13, 0.55, 0.13}

\biboptions{numbers,sort&compress}

\usepackage{hyperref}
\hypersetup{
    colorlinks=true,
    linkcolor=red,
    citecolor=blue,
    urlcolor=magenta
}

\begin{document}
\AtBeginDocument{%
  \hypersetup{
    colorlinks=true,
    linkcolor=BrickRed,
    citecolor=MidnightBlue,
    urlcolor=RoyalBlue,
    menucolor=Orange,
    pdfborder={0 0 0}
  }%
}

\title{Chiral, parity-doublet, effective-Lagrangian mean-field theories for nuclear and astrophysical phenomenology}
\author{Ayon Mukherjee}
\address{Brainware University, Barasat, Kolkata 700124, West Bengal, India}

\begin{abstract}
    Chiral-parity (parity-doublet) effective Lagrangian models provide a symmetry-consistent and economical framework for describing baryons and their negative-parity partners within a linearly realized chiral symmetry. In contrast to the conventional linear sigma model, the parity-doublet formulation admits a chirally invariant mass term, $m_0$, which allows baryons to retain finite masses; even as the chiral condensate vanishes. Within this setup, hadronic matter can be treated consistently across vacuum, nuclear and dense astrophysical environments. This review presents a focused synthesis of the essential structures of parity-doublet Lagrangians; outlines their mean-field implementation for nuclear and stellar matter; and discusses recent phenomenological and lattice constraints on the chirally invariant mass. Particular attention is given to mirror versus naïve chiral assignments; the role of vector interactions in achieving nuclear saturation; and the implications of parity doubling for the equation-of-state of dense matter and neutron-star cooling. The review is concluded by highlighting open theoretical challenges and possible directions for extending these models beyond the mean-field approximation.
\end{abstract}

\maketitle

\section{Introduction}
\label{sec:intro}

\noindent
Understanding the generation of baryon masses and the behaviour of strongly interacting matter under extreme conditions remains a central challenge in nuclear and hadronic physics. Chiral symmetry and its spontaneous breaking form the theoretical foundation of low-energy hadron dynamics~\cite{Nambu:1961tp, Nambu:1961fr,PhysRev.166.1568,Weise:2010cp}. While the emergence of nearly massless pions as Nambu–Goldstone bosons is well established~\cite{GellMann:1960np, Weinberg:1972kfs,PhysRevLett.18.188}, the microscopic origin of baryon masses and the fate of chiral symmetry at high baryon density remain open problems~\cite{Fukushima:2003fw,Fukushima:2006uv,Fukushima:2008is,Fukushima:2008wg,Fukushima:2009dx,Fukushima:2014lfa,Fukushima:2004bj,Fukushima:2013rx,Stoecker:1986ci,Schaefer:2007pw,Ratti:2005jh,PhysRevD.70.054013}.

\noindent
In this context, effective hadronic frameworks have been developed to connect QCD symmetries with nuclear phenomenology. Conventional relativistic mean-field (RMF) models and chiral effective field theory (EFT) approaches have achieved considerable success in describing nuclear matter and finite nuclei, yet they typically attribute the entirety of the nucleon mass to spontaneous chiral symmetry breaking. Parity-doublet models provide an alternative, symmetry-consistent formulation in which baryon masses receive both chirally invariant and symmetry-breaking contributions. Within this framework, finite nucleon masses can persist even in the chirally restored phase, thereby offering a natural setting to investigate dense-matter phenomenology, neutron-star structure and the restoration of chiral symmetry.

\noindent
A particularly appealing realisation of this idea was introduced by DeTar and Kunihiro~\cite{PhysRevD.39.2805}, who proposed that the nucleon and its opposite-parity partner form a chiral doublet. In such \emph{parity-doublet} (or \emph{chiral, parity-doublet}) models, a chirally invariant mass term, \( m_0 \), is allowed without violating chiral symmetry, in contrast to the traditional linear sigma model where baryon masses vanish in the restored phase. Subsequent analyses~\cite{Nemoto:1998um,Jido:2001nt,Jido:2003cb,Gallas:2009qp,Gallas:2013ipa} established two possible symmetry assignments for the baryon doublet: the \emph{naïve} assignment; in which chiral transformations act identically on both parity partners; and the \emph{mirror} assignment; in which left- and right-handed components are interchanged. The latter permits a non-zero \( m_0 \), ensuring that a finite component of the baryon mass survives even in a chirally restored medium.

\noindent
This structural feature has motivated extensive exploration of parity-doublet models as effective theories of strongly interacting matter across a wide range of densities and temperatures. When implemented in the mean-field approximation~\cite{Papazoglou:1997uw,Papazoglou:1998vr,Boguta:1977xi,boguta1983saturating,Serot:1984ey}, the resulting chiral-parity Lagrangian provides a tractable description of bulk nuclear matter and compact-star equations of state. Typical formulations couple the nucleon doublet to scalar (\(\sigma\)) and vector (\(\omega,\rho\)) mesons, successfully reproducing empirical saturation properties~\cite{Dexheimer:2008cv,Steinheimer:2010ib,Dexheimer:2012eu} over a broad range of \( m_0 \) values~\cite{Motohiro:2015taa,Minamikawa:2021fln}. Connections to lattice-QCD results, including indications of parity doubling near the chiral crossover~\cite{Aarts:2015mma,Yamazaki:2019tuo,cheng2009baryon, Gallas:2013ipa,Mukherjee:2016nhb,Steinheimer:2018rnd,Steinheimer:2010sp}, have further strengthened the phenomenological relevance of this framework. Applications have also been extended to hyperonic matter~\cite{Abuki:2008nm,Abuki:2008tx,Abuki:2008ht,Abu-Shady2017a,Mukherjee:2018yft,Adhikari2018}, neutron-star modelling~\cite{Mukherjee:2017jzi,Kong:2024review,Hanauske:2017oxo,Schramm2018,schramm2018dense,Schramm:2019npn,Schramm:2019oub,Steinheimer:2018abr,Marczenko:2018jui,Marczenko:2022jhl} and finite-temperature QCD phenomenology~\cite{Weyrich:2015hha}.

\noindent
At the same time, recent developments have clarified the broader theoretical context in which parity-doublet constructions are embedded. A comprehensive review of chiral symmetry breaking and hadron masses \cite{lee2023chiral} provides a conceptual bridge to these models, while quark-model analyses \cite{nefediev2025quark} supply historical perspective on hadron structure. The persistence of mass in QCD-like theories \cite{xu2026to} and studies of chiral potentials and Nambu–Goldstone modes \cite{moinester2025tests} further illuminate the underlying mechanisms. In addition, investigations of strong decays and spin-parity assignments of low-lying charmed baryons \cite{jakhad2024strong} highlight the broader challenges associated with identifying the nucleon’s parity partner. Collectively, these works situate parity-doublet models within the wider landscape of QCD-based effective theories.

\noindent
Despite these advances, a focused synthesis of theoretical structures, mean-field formulations and phenomenological constraints has remained lacking. This review aims to fill that gap by presenting a concise and coherent account of parity-doublet effective theories, with emphasis on key conceptual elements and selected recent developments rather than an exhaustive survey of all chiral approaches. Section~\ref{sec:framework} outlines the construction of the parity-doublet Lagrangian and its mean-field reduction. Section~\ref{sec:phenomenology} surveys phenomenological applications, including saturation properties and astrophysical implications. Section~\ref{sec:issues} discusses open theoretical issues and future directions, with particular emphasis on beyond–mean-field treatments and improved lattice constraints on \( m_0 \).

\section{Theory}
\label{sec:framework}

\subsection{Parity-doublet structure and chiral assignments}

\noindent
The central idea of the chiral-parity (parity-doublet) models is that the nucleon and its opposite-parity partner, denoted \( N^{+} \) and \( N^{-} \), are combined into a single multiplet of the chiral group \( SU(2)_{L} \times SU(2)_{R} \)~\cite{Casalbuoni1984,Hatsuda:1988mv}; as shown in Figure \ref{fig:chiral_parity_schematic}. Following DeTar and Kunihiro~\cite{PhysRevD.39.2805}, one introduces two independent Dirac fields, \( \psi_{1} \) and \( \psi_{2} \), which transform differently under chiral rotations.

\begin{figure}[h!]
\centering
\begin{tikzpicture}[node distance=2cm, every node/.style={align=center}]
    \node (chiral) [rectangle, draw, rounded corners, fill=white!20, minimum width=3cm, minimum height=1cm] {Chiral Symmetry\\ $SU(2)_L \times SU(2)_R$};
    
    \node (nucleon) [rectangle, draw, rounded corners, fill=white!20, below left=of chiral, xshift=-1cm] {Nucleon \\ $(N^+)$};
    \node (parity) [rectangle, draw, rounded corners, fill=white!20, below right=of chiral, xshift=1cm] {Parity Partner \\ $(N^-)$};
    
    \node (effective) [rectangle, draw, rounded corners, fill=white!20, below=of $(nucleon)!0.5!(parity)$] {Effective Mass \\ $M^* = m_0 + \Delta m(\sigma)$};
    
    \draw[->, thick] (chiral) -- (nucleon) node[midway, left] {chiral assignment~~};
    \draw[->, thick] (chiral) -- (parity) node[midway, right] {~~chiral assignment};
    \draw[->, thick] (nucleon) -- (effective);
    \draw[->, thick] (parity) -- (effective);
\end{tikzpicture}
\caption{Schematic linking chiral symmetry, parity partners and effective masses in the parity-doublet model.}
\label{fig:chiral_parity_schematic}
\end{figure}

\begin{figure}
    \centering
     \includegraphics[width=0.5\linewidth]{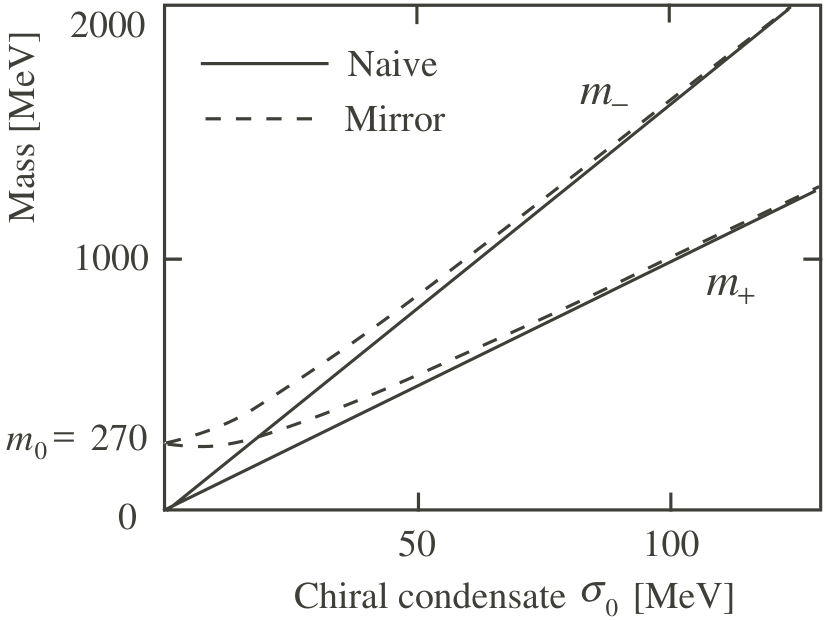}
    \caption{Masses of the positive- and negative-parity nucleons in the mirror and naïve assignments, as functions of the chiral condensate $\sigma_0$. Figure adopted from Ref.~\cite{hosaka2003progress}.}
    \label{fig:naive_vs_mirror}
\end{figure}

\paragraph{Mirror assignment}
In the \emph{mirror} assignment---now the standard version in mean-field studies---the fields transform as
\begin{align}
\psi_{1R} &\to R\,\psi_{1R}, & \psi_{1L} &\to L\,\psi_{1L}, \nonumber\\
\psi_{2R} &\to L\,\psi_{2R}, & \psi_{2L} &\to R\,\psi_{2L},
\label{eq:mirror-transf}
\end{align}
so that $\psi_{2}$ transforms oppositely to $\psi_{1}$. A chirally invariant mass term can then be constructed as
\begin{equation}
\mathcal{L}_{m_0} = -m_{0}\,\left(\overline\psi_{1}\gamma_{5}\psi_{2} - \overline\psi_{2}\gamma_{5}\psi_{1}\right),
\label{eq:Lm0}
\end{equation}
where \( m_{0} \) is a constant mass parameter that survives even when the chiral condensate vanishes. The Yukawa couplings to the scalar \( \sigma \) and pseudoscalar \( \vec{\pi} \) fields are
\begin{equation}
\mathcal{L}_{Y} =
- g_{1}\,\overline\psi_{1}\left(\sigma + i\gamma_{5}\vec{\tau}\!\cdot\!\vec{\pi}\right)\psi_{1}
- g_{2}\,\overline\psi_{2}\left(\sigma - i\gamma_{5}\vec{\tau}\!\cdot\!\vec{\pi}\right)\psi_{2}.
\label{eq:LY}
\end{equation}
Combining Equations~(\ref{eq:Lm0}) and~(\ref{eq:LY}) yields the total baryonic part of the Lagrangian,
\begin{equation}
\mathcal{L}_{B} = \overline\psi_{1} i\slashed{\partial} \psi_{1} +
\overline\psi_{2} i\slashed{\partial} \psi_{2} + \mathcal{L}_{Y} + \mathcal{L}_{m_0}.
\label{eq:LB}
\end{equation}
After spontaneous chiral symmetry breaking with \( \langle\sigma\rangle = \sigma_{0} \neq 0 \), the mass matrix in the $(\psi_{1},\psi_{2})$ basis reads
\[
\mathcal{M} = 
\begin{pmatrix}
g_{1}\sigma_{0} & m_{0}\gamma_{5}\\[2pt]
- m_{0}\gamma_{5} & g_{2}\sigma_{0}
\end{pmatrix}.
\]
Diagonalizing this matrix by a chiral rotation leads to physical fields \( N^{+} \) and \( N^{-} \) with masses
\begin{equation}
m_{N^{\pm}} = \frac{1}{2}\left[
\sqrt{(g_{1}+g_{2})^{2}\sigma_{0}^{2} + 4m_{0}^{2}}
\mp (g_{1}-g_{2})\,\sigma_{0}
\right].
\label{eq:mass-formula}
\end{equation}
Equation~(\ref{eq:mass-formula}) encapsulates the defining feature of the model: even when $\sigma_{0}\!\to\!0$ (chiral restoration),
the baryons retain a finite chirally invariant, mass \(m_{0}\); as evidenced by Figure \ref{fig:naive_vs_mirror}.

\paragraph{Naïve assignment}
For comparison, the \emph{naïve} assignment applies identical chiral transformations to both fields, which forbids the \(m_{0}\) term.
Masses then arise solely from the Yukawa coupling to the condensate: \( m_{N^{\pm}} = g_{\pm}\sigma_{0} \). This distinction strongly influences the pattern of chiral restoration and the behaviour of the equation-of-state (EoS) at high density~\cite{Jido:2003cb,Gallas:2013ipa}.

\subsection{Inclusion of vector and scalar mesons}

\noindent
To describe nuclear matter realistically, scalar and vector mesons are incorporated in analogy with RMF models~\cite{Walecka:1974qa,Bender:2003jk,boguta1983saturating}. A minimal Lagrangian for the mesonic sector is
\begin{equation}
\mathcal{L}_{M} =
\tfrac{1}{2}\left(\partial_{\mu}\sigma\right)^{2}
+ \tfrac{1}{2}\left(\partial_{\mu}\vec{\pi}\right)^{2}
- U\left(\sigma,\vec{\pi}\right)
- \tfrac{1}{4}\omega_{\mu\nu}\omega^{\mu\nu}
+ \tfrac{1}{2}m_{\omega}^{2}\omega_{\mu}\omega^{\mu}
- \tfrac{1}{4}\vec{\rho}_{\mu\nu}\!\cdot\!\vec{\rho}^{\,\mu\nu}
+ \tfrac{1}{2}m_{\rho}^{2}\vec{\rho}_{\mu}\!\cdot\!\vec{\rho}^{\,\mu},
\label{eq:meson}
\end{equation}
where \( U(\sigma,\vec{\pi}) \) is the chiral potential, typically of the Mexican-hat form \( U\left(\sigma,\vec{\pi}\right) = \frac{\lambda}{4}\left(\sigma^{2}+\vec{\pi}^{2}-f_{\pi}^{2}\right)^{2} - h\sigma \). The baryons couple to vector fields through
\begin{equation}
\mathcal{L}_{\mathrm{int}}^{V} =
- g_{\omega}\,\overline{N}\gamma_{\mu}\omega^{\mu} N
- g_{\rho}\,\overline{N}\gamma_{\mu}\vec{\tau}\!\cdot\!\vec{\rho}^{\,\mu} N,
\label{eq:LV}
\end{equation}
where \( N = (N^{+}, N^{-}) \). Such vector interactions are essential for nuclear saturation and for reproducing empirical incompressibility~\cite{Motohiro:2015taa,Minamikawa:2021fln,Cleymans:1992jz}. Extensions to three flavours and to hyperons introduce additional scalar and vector fields (\( \zeta,\phi \)), but their structure parallels Equation~(\ref{eq:LV})~\cite{Mukherjee:2017jzi}.

\subsection{Mean-field approximation}

\noindent
At the mean-field level, mesonic fields are replaced by their expectation values~\cite{Aarts:2017rrl,Berges:2000ew,Fodor:2004nz,Fukushima:2009dx,Gallas:2009qp,Herold:2016uvv,Rafelski:1982pu,Roessner:2006xn,Steinheimer:2010ib,Veneziano1982}:
\begin{equation}
\sigma \to \overline{\sigma}, \quad
\omega_{\mu} \to \delta_{\mu0}\,\overline{\omega}, \quad
\rho_{3}^{\mu} \to \delta_{\mu0}\,\overline{\rho}~,
\label{eq:MF-replace}
\end{equation}
while pion mean fields vanish in the mean-field ground state due to parity conservation of the strong interaction. The single-particle energies for baryons of species \(i = \{N^{+}, N^{-}\}\)
are
\begin{equation}
E_{i}(k) = \sqrt{k^{2} + {m^{*}_{i}}^{2}} + g_{\omega}\overline{\omega} + t_{3i}g_{\rho}\overline{\rho}~,
\label{eq:Ei}
\end{equation}
where the effective masses \(m^{*}_{i}\) follow from Equation~(\ref{eq:mass-formula}) with \(\sigma_{0}\!\to\!\overline{\sigma}\). The grand-canonical potential density at temperature \(T\) and chemical potential \(\mu\) is
\begin{equation}
\begin{aligned}
\Omega(T,\mu;\overline{\sigma},\overline{\omega},\overline{\rho}) = 
&- \sum_{i=N^{+},N^{-}} \frac{\gamma_i}{(2\pi)^3} \int d^3 k \,
    \Bigl[ T \ln\Bigl( 1 + e^{-\left(E_i(k)-\mu_i\right)/T} \Bigr)
          + T \ln\Bigl( 1 + e^{-\left(E_i(k)+\mu_i\right)/T} \Bigr) \Bigr] \\
&+ U(\overline{\sigma})
 - \frac{1}{2} m_\omega^2 \, \overline{\omega}^2
 - \frac{1}{2} m_\rho^2 \, \overline{\rho}^2~;
\end{aligned}
\label{eq:Omega}
\end{equation}
where \( \gamma_{i}(=4) \) is the spin–isospin degeneracy. In the zero-temperature limit, the logarithmic terms reduce to the usual Fermi-sea contributions.

\noindent
The mean-field (gap) equations follow from the stationarity conditions
\begin{equation}
\frac{\partial \Omega}{\partial \overline{\sigma}} = 0~, \qquad
\frac{\partial \Omega}{\partial \overline{\omega}} = 0~, \qquad
\frac{\partial \Omega}{\partial \overline{\rho}} = 0~,
\label{eq:gap}
\end{equation}
which are solved self-consistently at given baryon density. The pressure and energy density are obtained as \( P = -\Omega \) and \( \varepsilon = \Omega + \sum_{i}\mu_{i}\rho_{i} \). Nuclear saturation properties are fitted by adjusting \( (g_{1},g_{2},g_{\omega},m_{0}) \) and the parameters of \(U(\sigma)\).
For realistic sets, one obtains saturation at \( \rho_{0}\!\simeq\!0.16~\mathrm{fm}^{-3} \), binding energy \( B/A \!\simeq\! -16~\mathrm{MeV} \) and incompressibility \( K \!\sim\! 200\text{--}300~\mathrm{MeV} \)~\cite{Motohiro:2015taa}.
The chiral-invariant mass, \(m_{0}\), typically lies in the range \(500\text{--}900~\mathrm{MeV}\), consistent with phenomenological constraints~\cite{Yamazaki:2019tuo,Minamikawa:2021fln}.

\subsection{Beyond mean field and renormalization effects}

\noindent
Although the mean-field approximation captures the gross features of nuclear matter, it neglects fluctuations that can modify the chiral transition pattern. Functional-renormalization-group (FRG) studies~\cite{Weyrich:2015hha} show that mesonic and baryonic loop corrections soften the effective potential and can generate non-trivial correlations between \( \sigma \) and \( \omega \) fields~\cite{Ferreira:2013tba,Satarov:2009zx,Rossner:2007ik}. Such beyond-mean-field effects become particularly relevant near the chiral critical point~\cite{Stephanov:1998dy}, where the simple mean-field order parameter evolution may be qualitatively altered. Recent analyses further attempt to renormalize the parity-doublet model in a way consistent with vector-meson dominance and hidden local symmetry, providing a bridge between the effective hadronic and holographic descriptions~\cite{Kong:2024review}.

\subsection{Summary of theoretical framework}

\noindent
The chiral-parity (parity-doublet) model, thus, unifies several appealing features; as mentioned below:
\begin{enumerate}
\item it realizes linear chiral symmetry for baryons while allowing a finite invariant mass \(m_{0}\);
\item it reproduces empirical nuclear saturation via vector interactions;
\item it provides a natural platform for exploring chiral restoration and parity doubling in dense matter; and
\item it remains extensible to finite temperature, three flavours and astrophysical conditions.
\end{enumerate}
These theoretical foundations constitute the basis for the phenomenological studies summarized in Section~\ref{sec:phenomenology}.

\section{Phenomenology}
\label{sec:phenomenology}

\noindent
The chiral-parity (parity-doublet) mean-field framework, although conceptually simple, has proved flexible enough to describe a broad range of nuclear and astrophysical observables.  This section summarizes key phenomenological results, beginning with nuclear-matter saturation and extending to asymmetric systems, neutron stars and the connection to lattice-QCD constraints.

\begin{figure}[H]
    \centering
     \includegraphics[width=0.5\linewidth]{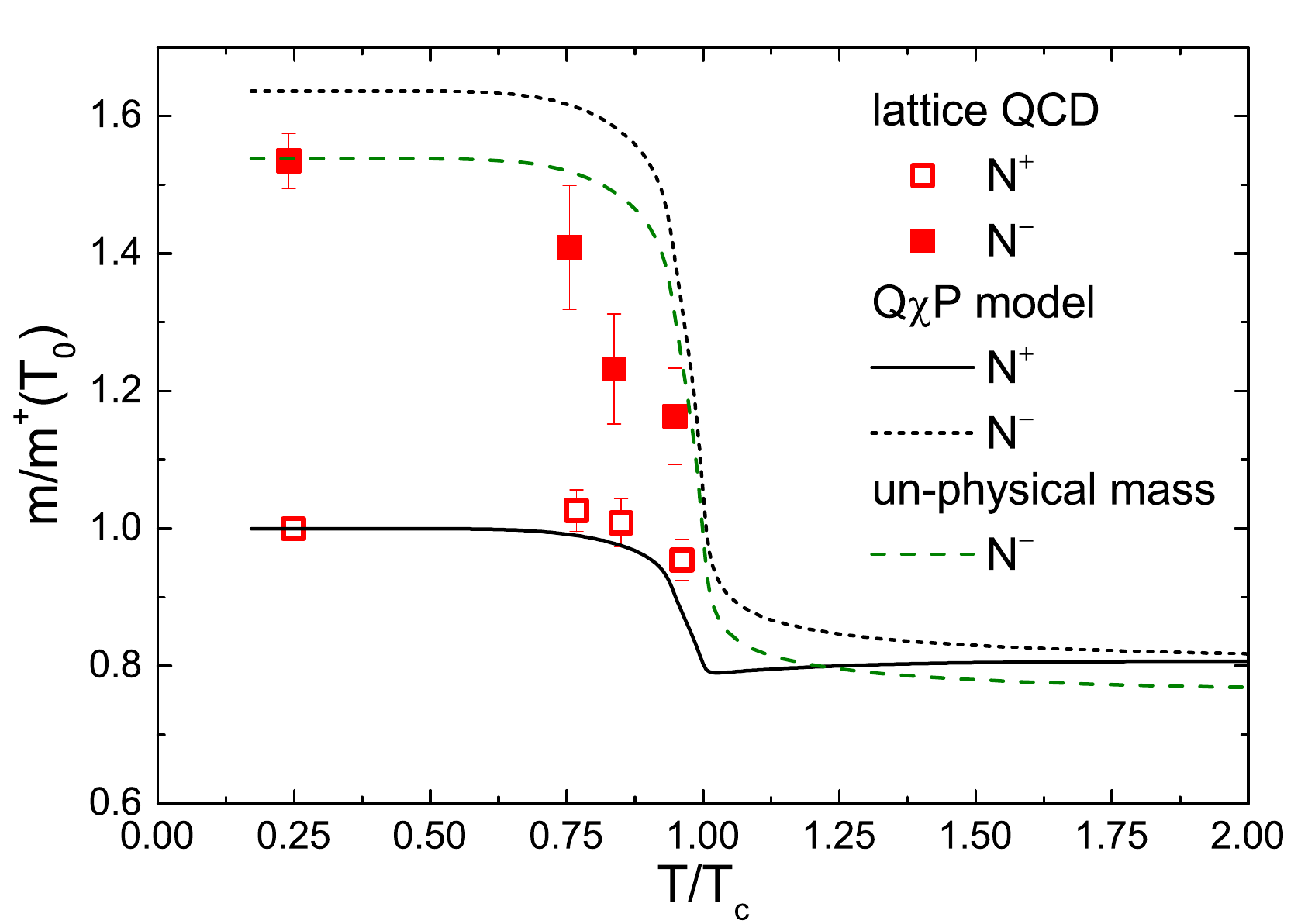}
    \caption{Temperature dependence of the effective masses of the positive- and negative-parity nucleons in one of the parity-doublet mean-field models. The convergence of \( m_{N^+}^* \) and \( m_{N^-}^* \) at high \( T \) reflects chiral restoration.}
    \label{fig:mass-vs-T}
\end{figure}

\subsection{Nuclear matter saturation and compressibility}
\label{subsec:saturation}

\noindent
At the mean-field level, bulk nuclear matter properties are determined by minimizing the thermodynamic potential with respect to the expectation values of the meson fields~\cite{Serot:1984ey,Cohen:1991nk,RUSNAK1997495,FURNSTAHL1997441}. The inclusion of $m_0$ has a crucial effect on the saturation mechanism~\cite{Motohiro:2015taa,PhysRevC.75.055202}. In the standard Walecka model, saturation arises from the competition between large scalar attraction and vector repulsion~\cite{Walecka:1974qa,Rischke:1991ke}. In the parity-doublet formulation, however, the scalar field couples not to the entire baryon mass but only to the chiral symmetry breaking component, effectively weakening the scalar response and modifying the density dependence of the nucleon effective masses~\cite{Motohiro:2015taa}. This allows for realistic binding even with comparatively modest scalar fields~\cite{PhysRevC.75.055202}.

\noindent
Motohiro, Kim and Harada~\cite{Motohiro:2015taa} performed a systematic fit of the mean-field parameters to reproduce the empirical saturation density \( \rho_0 \simeq 0.16~\mathrm{fm^{-3}} \), binding energy per nucleon \( E/A \simeq -16~\mathrm{MeV} \) and symmetry energy \( a_4 \simeq 31~\mathrm{MeV} \).  
They found that successful fits require a chirally invariant mass in the range \( m_0 \simeq 500\!-\!900~\mathrm{MeV} \), with \( m_0 \sim 700~\mathrm{MeV} \) yielding optimal compressibility \( K \simeq 250~\mathrm{MeV} \) and effective mass ratio \( M_N^*/M_N \approx 0.7 \) at saturation.  
This parameter window has since been widely employed in phenomenological studies.

\noindent
The parity-doublet structure introduces a smooth crossover from vacuum to to hot hadronic matter, where the positive-parity nucleon mass decreases with increasing temperature the negative-parity partner (\( N^- \)) mass approaches degeneracy.  
The resulting behaviour is illustrated in Figure~\ref{fig:mass-vs-T}, which shows the temperature dependence of the effective masses \( m_{N^\pm}^*(T) \).  
At low temperature, \( N^- \) remains heavy and decoupled, but as the chiral condensate melts, the masses converge toward \( m_0 \), signifying chiral restoration.  

\noindent
The value of \( m_0 \) controls the relative stiffness of the nuclear EoS.  
Smaller \( m_0 \) leads to stronger scalar attraction and, consequently, a softer EoS; while larger \( m_0 \) yields a stiffer response comparable to conventional RMF parameterizations such as NL3 or TM1.  
This tunability has been exploited to map EoS uncertainties relevant for both nuclear and astrophysical observables~\cite{Minamikawa:2021fln}.

\subsection{Finite nuclei and asymmetric nuclear matter}
\label{subsec:finite-nuclei}

\noindent
Extensions of the parity-doublet model to finite nuclei have been developed to examine the microscopic consistency of the mean-field framework. By incorporating gradient terms together with Coulomb corrections, binding energies and charge radii across medium-mass nuclei can be reproduced at a level comparable to standard RMF approaches~\cite{Sasaki:2006ww,Kong:2024review}. Owing to the mirror assignment, the negative-parity partner contributes only minimally at normal nuclear density, such that an effective single-nucleon structure closely resembling the RMF description is retained, while remaining grounded in chiral symmetry.

\noindent
Beyond purely hadronic descriptions, parity-doublet models have also been embedded in hybrid frameworks that permit a transition or crossover to quark matter at high density. A representative example is the coupling of the parity-doublet model to a Nambu–Jona-Lasinio–type quark sector, through which first-order phase transitions in neutron-star interiors can be investigated. Such studies indicate that the emergence and strength of a phase transition depend sensitively on the value of $m_0$ and on the adopted quark–hadron matching procedure, thereby illustrating that constraints on the chirally invariant mass are closely intertwined with assumptions regarding the microscopic composition of ultra-dense matter \cite{gao2024exploring}.

\noindent
In asymmetric matter, the interplay between scalar-isovector (\(\delta\)) and vector-isovector (\(\rho\)) meson couplings determines the density dependence of the symmetry energy. Parity-doublet models incorporating both channels~\cite{Kong:2023a0} are found to reproduce empirical symmetry-energy coefficients and slope parameters \( L \sim 50\!-\!70~\mathrm{MeV} \), consistent with constraints from isobaric analogue states and heavy-ion collisions~\cite{Reisdorf:2010ie}. These features are essential for predictions of the neutron-proton mass splitting in neutron-rich matter and for establishing a link between terrestrial nuclear data and astrophysical observables.

\subsection{Neutron stars and cooling phenomenology}
\label{subsec:neutron-stars}

\noindent
One of the strongest motivations for parity-doublet models arises from their implications for the EoS of dense matter and the structure of compact stars~\cite{PhysRev.55.364}. The value of \( m_0 \) governs the rate at which the nucleon effective masses decrease with increasing density. This behaviour directly influences the pressure–density relation and, consequently, the maximum neutron-star mass attainable within the model~\cite{Baym:1971pw}. Related correlations between nuclear and neutron-star radii have also been examined in Ref.~\cite{Schramm:2002xa}.

\begin{figure}[htb]
  \centering
  \begin{subfigure}[b]{0.45\linewidth}
      \centering
      \includegraphics[angle=270,width=\linewidth]{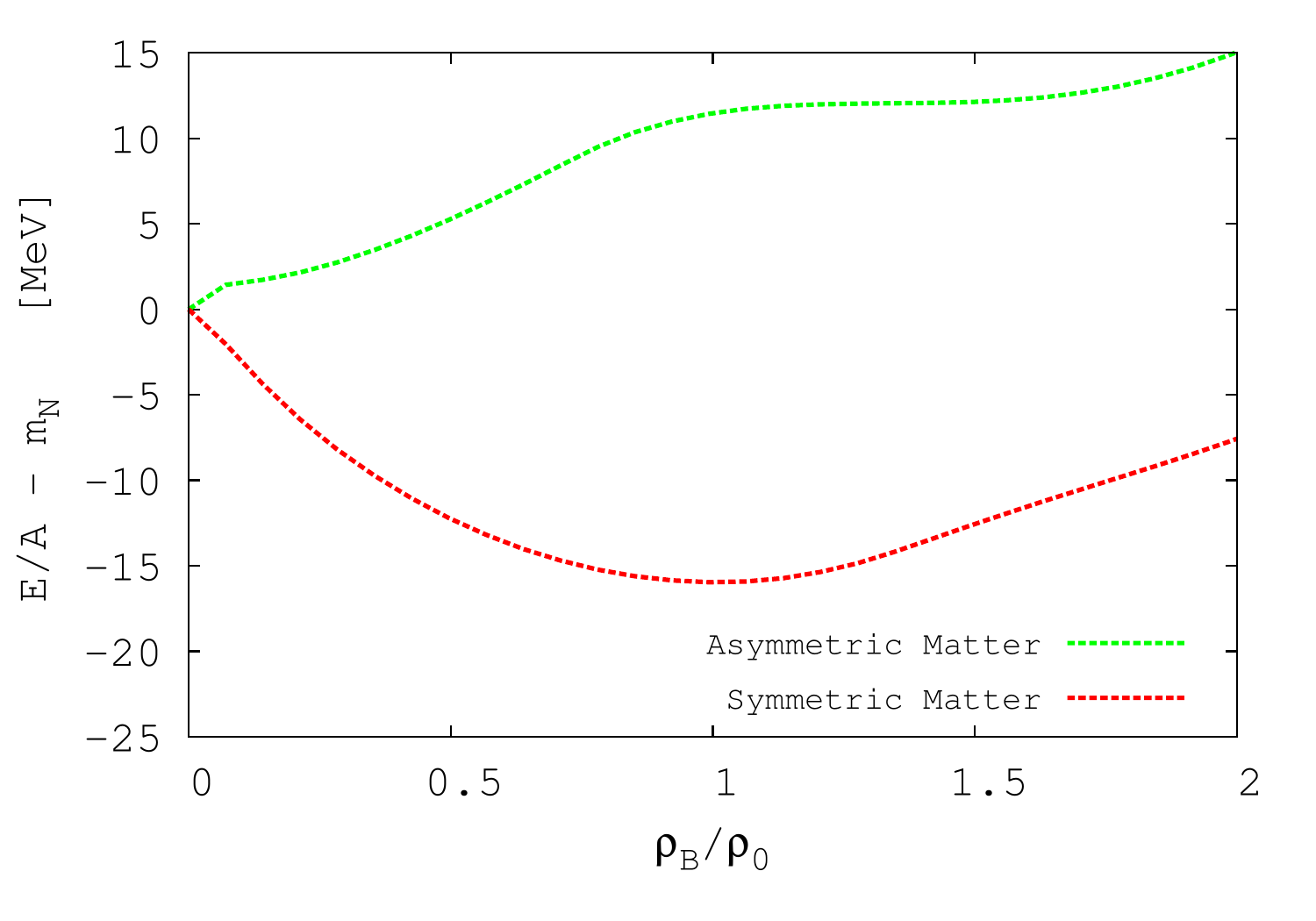}
      \caption{Parity-doublet mean-field model}
      \label{fig:eos-comparison-a}
  \end{subfigure}
  \hfill
  \begin{subfigure}[b]{0.45\linewidth}
      \centering
      \includegraphics[width=\linewidth]{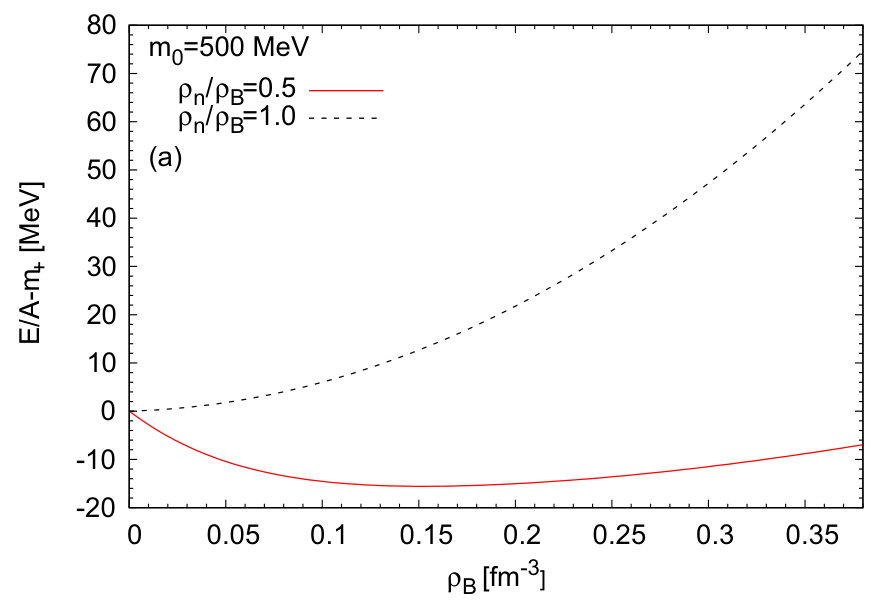}
      \caption{Standard RMF (NL3 parameter set)}
      \label{fig:eos-comparison-b}
  \end{subfigure}
  \caption{Binding energy comparison: (a) binding energy $\left(E/A - m_N\right)$ as a function of normalised baryon density $\left(\rho_B/\rho_0\right)$ for one of the parity-doublet mean-field models and (b) the corresponding standard RMF (NL3) result, adopted from Ref.~\cite{Motohiro:2015taa}. The parity-doublet EoS softens at moderate densities and stiffens at higher densities due to the density dependence of the effective masses \( m_{N^\pm}^* \).}
  \label{fig:eos-comparison}
\end{figure}

\noindent
Calculations by Motohiro \textit{et al.}~\cite{Motohiro:2015taa}, followed by refinements from Minamikawa \textit{et al.}~\cite{Minamikawa:2021fln}; Mukherjee \textit{et al.}~\cite{Mukherjee:2017jzi}; and Dexheimer \textit{et al.}~\cite{Dexheimer:2007tn}, have demonstrated that parity-doublet equations of state can support neutron-star masses in the range \( 2.0\!-\!2.2\,M_\odot \), consistent with recent pulsar observations~\cite{Antoniadis1233232,Steiner:2010fz}, while remaining compatible with empirical nuclear-matter constraints~\cite{dexheimer2015reconciling}. Figure~\ref{fig:eos-comparison} presents a comparison between a representative parity-doublet EoS and a standard RMF curve (NL3 parameter set) adopted from Ref.~\cite{Motohiro:2015taa}. The parity-doublet EoS is generally softer at intermediate densities but becomes stiffer beyond \( \sim 3\rho_0 \) as partial chiral symmetry restoration sets in, yielding comparable maximum masses but distinct internal compositions and proton fractions.

\noindent
As the density increases, the parity partner \( N^- \) is gradually populated, leading to modifications in the proton fraction and in the threshold for the direct Urca (DU) process relative to conventional RMF models. Recent analyses~\cite{Kong:2024review,Blaschke:2005dc} suggest that parity-doubled stars may cool more rapidly, owing to enhanced DU processes and modified neutrino-emission rates mediated through the \( N^- \) channel~\cite{Page:2005fq,Page:2004fy}. Such effects could provide observable signatures if correlated with surface-temperature measurements of young neutron stars.

\noindent
The parity-doublet EoS is also characterised by a smooth crossover rather than a strong first-order chiral transition. This feature ensures thermodynamic stability and facilitates unified descriptions of hadronic and quark degrees of freedom within a single effective framework~\cite{Ciminale:2007sr}. In conjunction with quark–hadron continuity scenarios, parity-doublet mean-field models may, therefore, be interpreted as effective low-energy realizations of QCD at finite density.

\noindent
In recent years, neutron-star observations have increasingly been employed to constrain the value of $m_0$ within extended parity-doublet frameworks. In particular, a series of studies by Gao, Harada and collaborators has combined parity-doublet descriptions of dense hadronic matter with astrophysical data from massive pulsars and the supernova remnant HESS J1731-347. These investigations indicate that values of $m_0$ in the range $600\text{--}800$ MeV are favoured once simultaneous constraints from maximum neutron-star masses, radius measurements and tidal deformabilities are imposed. Extensions incorporating isovector scalar mesons further sharpen these bounds, underscoring the importance of isospin-dependent interactions in dense matter. Collectively, these analyses reinforce the view that astrophysical observations provide non-trivial and complementary constraints on the origin of nucleon mass in parity-doublet models \cite{kong2025chiral,gao2024recon,gao2025origin,gao2026imp}.

\subsection{Lattice-QCD and experimental constraints}
\label{subsec:lattice}

\noindent
Independent support for the parity-doublet picture is provided by lattice-QCD simulations, which observe parity doubling in the baryon spectrum near the chiral crossover~\cite{Borsanyi:2010bp,Aoki:2006we,Bazavov:2012jq,Alexandrou2018}, as well as by comprehensive lattice-QCD thermodynamic studies performed with physical quark masses~\cite{Soltz:2015ula}. Degeneracy of positive- and negative-parity correlators at temperatures above \( T_c \), implying approximate restoration of chiral symmetry, was first reported by Aarts \textit{et al.}~\cite{Aarts:2015mma}. This degeneracy was subsequently quantified by Yamazaki \textit{et al.}~\cite{Yamazaki:2019tuo}, who extracted the temperature dependence of the parity splitting and provided numerical evidence for a substantial chirally invariant mass component \( m_0 \sim 700~\mathrm{MeV} \).

\noindent
Complementary evidence has been obtained from QCD sum-rule analyses. In particular, the nucleon and its negative-parity partner were investigated in nuclear matter by Kim and Lee~\cite{kim2021vector}, where the in-medium behaviour of the two states was found to be consistent with a sizable chirally invariant mass component. The observed reduction of parity splitting with increasing density, while maintaining a non-vanishing mass in the chirally restored regime, was shown to be compatible with the parity-doublet interpretation.

\noindent
Taken together, these lattice and continuum-QCD results indicate a coherent picture across temperature and density scales, thereby providing additional theoretical support for effective descriptions that incorporate a chirally invariant mass term.

\noindent
Experimental indications for the framework are also found in the identification of parity-partner states in the baryon spectrum~\cite{Adamczyk:2014fia,PhysRevLett.112.032302,Aggarwal:2010wy}. The \( N(1535) \) resonance is commonly interpreted as the negative-parity partner of the nucleon, although alternative assignments remain under debate~\cite{Gallas:2013ipa}. Electroproduction and photoproduction data have provided constraints on axial coupling constants and transition form factors, which in turn have been used to constrain the parameters of effective parity-doublet Lagrangians~\cite{PhysRevC.68.055501,PhysRevC.83.025208,Anisovich2011,Kashevarov2009}.

\noindent
At finite density, heavy-ion experiments probing chiral restoration through dilepton spectra~\cite{Seck2022dilepton,Agakishiev:2009am} or possible vector-meson mass shifts may offer further tests of the framework.

\noindent
Despite these developments, important uncertainties remain. The precise determination of \( m_0 \) continues to exhibit model dependence and the coupling of the parity partners to vector mesons has not yet been tightly constrained. Moreover, beyond–mean-field effects, including pion loops, correlations and density fluctuations, may modify both the saturation and chiral-restoration behaviour. Ongoing efforts seek to combine the parity-doublet mean-field framework with functional renormalization group (FRG) or Dirac–Brueckner approaches in order to address these issues~\cite{Weyrich:2015hha}.

\noindent
In this way, parity-doublet effective Lagrangians are seen to provide a unified and symmetry-consistent description of hadronic matter from the vacuum to compact-star densities. By encoding both explicit and dynamical chiral symmetry breaking, a bridge is established between phenomenological RMF models and fundamental QCD expectations. The continued interplay between theory, lattice results and astrophysical observations will ultimately determine the extent to which this framework can serve as a realistic EFT of strongly interacting matter.

\subsection{Beyond conventional RMF}
\label{subsec:rmf}

\noindent
While parity-doublet mean-field models share many structural similarities with conventional RMF theories, their physical interpretation differs in a crucial way. In standard RMF models, the nucleon mass is entirely generated by scalar mean fields and chiral symmetry plays no explicit role. As a result, the density dependence of the effective mass and the stiffness of the EoS are largely controlled by phenomenological scalar self-interactions.

\noindent
In contrast, parity-doublet models introduce a clear separation between chirally invariant and symmetry-breaking contributions to the baryon mass. The presence of a non-zero $m_0$ weakens the direct coupling between the chiral order parameter and the bulk thermodynamics, leading to a qualitatively different interpretation of saturation and chiral restoration. In particular, the reduction of the scalar condensate does not imply the vanishing of baryon masses, which allows for a smoother crossover to chirally restored matter.

\noindent
This structural difference has observable consequences. For a fixed saturation point, parity-doublet models generally predict a milder decrease of the nucleon effective mass with density and a delayed onset of strong chiral restoration compared to RMF models. At the same time, this advantage comes at the cost of increased parameter sensitivity: the stiffness of the high-density EoS is strongly correlated with the choice of $m_0$ and different parameter sets can lead to quantitatively distinct neutron-star predictions. Thus, while parity doubling provides a symmetry-motivated extension of RMF theory, its predictive power ultimately depends on independent constraints on the invariant mass and vector couplings.

\section{Issues}
\label{sec:issues}

\noindent
Despite substantial progress, several conceptual and phenomenological challenges continue to limit the establishment of the chiral-parity mean-field framework as a quantitatively predictive theory of hadronic matter. These open issues concern both the microscopic identification of the relevant degrees of freedom and the systematic incorporation of physics beyond the mean-field approximation.

\noindent
From a critical standpoint, the principal strength of parity-doublet mean-field models resides in their symmetry structure rather than in demonstrable quantitative novelty. By embedding nuclear saturation and dense-matter phenomenology within a linearly realized chiral framework, a coherent interpretation of baryon mass generation is provided that is absent in purely phenomenological RMF approaches. However, this conceptual advantage does not automatically translate into enhanced predictive power unless $m_0$ and the vector couplings are independently constrained. 

\noindent
At a more formal level, renormalization-group invariant formulations of the parity-doublet model have recently been proposed in order to reduce scheme dependence and improve theoretical consistency. Such developments may prove essential if quantitative constraints on $m_0$ are to be rendered robust against variations in the effective description.

\subsection{Identification of the nucleon's parity partner}

\noindent
A first unresolved issue concerns the identification of the negative-parity partner of the nucleon. In most phenomenological implementations, the \( N(1535) \) resonance is assumed to be the chiral partner of the ground-state nucleon, primarily because of its matching quantum numbers and comparatively low mass. However, analyses by Jido \textit{et al.}~\cite{Jido:2003cb} and Gallas \textit{et al.}~\cite{Gallas:2013ipa} have indicated that this identification may be overly simplistic. Coupled-channel studies of the \( N(1535) \) suggest a substantial meson–baryon molecular component rather than a simple three-quark structure, while other negative-parity states such as \( N(1650) \) or mixed configurations may constitute a more appropriate chiral partner. The uncertainty in the partner assignment propagates directly into the determination of \( m_0 \), since different choices yield distinct fits to vacuum spectra and in-medium properties. A consistent identification will likely require combined input from lattice spectroscopy; coupled-channel analyses; and electroproduction data on axial and transition form factors.

\subsection{Scale-symmetry and the origin of $m_0$}

\noindent
Beyond phenomenological constraints, a more fundamental question concerns the microscopic origin of $m_0$. While the parity-doublet model parametrizes it as a constant, its dynamical origin remains unresolved. Since $m_0$ does not vanish when chiral symmetry is restored, it cannot originate from the quark condensate. This naturally raises the possibility that it is instead related to the breaking of scale symmetry in QCD, namely to the trace anomaly.

\noindent
In QCD, the nucleon mass may be decomposed through the matrix element of the energy–momentum tensor, whose trace receives contributions from the gluonic trace anomaly. Recent studies of gravitational form factors have emphasized that a substantial fraction of the nucleon mass originates from gluonic dynamics and is not directly tied to chiral symmetry breaking. Such analyses suggest that the chirally invariant component of the nucleon mass may be associated with gluon condensates and the non-perturbative structure of the QCD vacuum \cite{Kawaguchi2025cuf,Stegeman2025sca,Stegeman2025tdl}.

\noindent
Within effective approaches, scale symmetry is frequently implemented via a dilaton field that mimics the trace anomaly. Several works have explored the possibility that $m_0$ is dynamically generated through coupling to a dilaton condensate, thereby linking the chirally invariant mass to the behaviour of scale symmetry in medium. In such scenarios, partial restoration of scale symmetry at high density could induce a density dependence in $m_0$, which would in turn affect neutron-star phenomenology \cite{rho2017proton,paeng2013inteplay,Ma2019pseudo,MA2020progress}.

\noindent
Although a definitive identification of $m_0$ with a specific QCD operator remains elusive, these developments point towards a deeper interpretation of the chirally invariant mass beyond that of a purely phenomenological parameter. Establishing a quantitative bridge between parity-doublet models and QCD-based decompositions of the nucleon mass therefore constitutes an important direction for future investigation.

\subsection{Beyond–mean-field effects}

\noindent
In addition to questions regarding microscopic identification, an important limitation of the current framework lies in its reliance on the mean-field approximation. Although mean-field treatments efficiently capture bulk thermodynamics and nuclear saturation, mesonic fluctuations, correlations and collective excitations are neglected, even though they become essential near the chiral transition~\cite{Chomaz:2003dz}.  
Weyrich, Tripolt and Wambach~\cite{Weyrich:2015hha,Schaefer:2011ex} demonstrated, within a functional renormalization-group (FRG) extension, that fluctuations can substantially modify both the order and the location of the chiral transition. In particular, the inclusion of pionic and sigma modes smooths the transition and shifts the pseudocritical temperature relative to the mean-field prediction. At finite density, such effects may alter the effective mass evolution and soften the EoS around \( 2\!-\!3\rho_0 \), thereby influencing neutron-star radii and cooling behaviour. Future theoretical developments should therefore aim at embedding the parity-doublet Lagrangian within fluctuation-sensitive frameworks such as FRG, Dyson–Schwinger or density-functional renormalization methods.

\subsection{Constraining $m_0$ and extending the framework}

\noindent
Closely connected to these formal developments is the determination of \( m_0 \), which remains a pivotal parameter of the framework. Its present empirical range, \( m_0 \approx 500\!-\!900~\mathrm{MeV} \), is inferred indirectly from nuclear matter and neutron-star data~\cite{Motohiro:2015taa,Minamikawa:2021fln}. However, lattice-QCD studies of parity-doubling at finite temperature~\cite{Aarts:2015mma,Yamazaki:2019tuo,Borsanyi:2013bia} suggest \( m_0 \simeq 700~\mathrm{MeV} \) with substantial systematic uncertainties~\cite{Endrodi:2011gv,Fodor:2004nz}. A major task ahead will be to perform a global Bayesian analysis combining lattice correlators, nuclear saturation constraints and astrophysical observables (mass–radius and tidal deformability) in order to extract \( m_0 \) and its correlations with other couplings. Recent reviews~\cite{Kong:2024review} have emphasized that such multi-channel analyses could elevate the parity-doublet model from a qualitative concept to a precision EFT for dense QCD.

\noindent
A further frontier lies in extending the model to the three-flavour sector, incorporating the \( \Lambda \), \( \Sigma \) and \( \Xi \) baryons together with their parity partners. Such a generalization is essential for studying the appearance of hyperons in neutron-star cores and for exploring the interplay between chiral restoration and strangeness. Three-flavour parity-doublet models~\cite{schulze1998hyperonic,Sasaki:2010bp,Steinheimer:2011ea,Mukherjee:2018yft,Tsubakihara:2009zb} predict that hyperonic parity-doubling may delay the onset of hyperons, thereby mitigating the conventional hyperon-softening problem. They also provide a unified platform for connecting heavy-ion-collision observables, such as fluctuations of conserved charges~\cite{Borsanyi:2011sw,Borsanyi:2013hza}, with astrophysical EoS constraints.

\subsection{Phenomenological and observational interfaces}

\noindent
The interface between theory and experiment has expanded rapidly in recent years. Heavy-ion experiments at RHIC, FAIR and NICA probe baryon densities up to several times saturation and temperatures near the chiral crossover~\cite{Chen:2011am,Chen:2016sxn,Schaefer:2009ui}. Parity-doublet models are capable of providing testable predictions for the in-medium spectral behaviour of parity partners~\cite{Steinheimer:2018rnd}, dilepton yields~\cite{Seck2022dilepton} and signatures of chiral restoration~\cite{Mukherjee:2016nhb,Mukherjee:2018yft}. In astrophysics, the advent of gravitational-wave measurements (GW150914~\cite{PhysRevLett.116.061102,2041-8205-818-2-L22}, GW170817~\cite{PhysRevLett.113.091104}, GW190425~\cite{Abbott_2020}) and precision X-ray timing from NICER has provided new constraints on the neutron-star EoS that directly affect the allowed range of \( m_0 \) and vector couplings~\cite{PhysRevD.93.124051}. Consistency between these constraints and parity-doublet model predictions will constitute a decisive test of the framework.

\noindent
Thus, chiral-parity mean-field models have matured into a versatile bridge between low-energy nuclear phenomenology and QCD thermodynamics. Their full potential, however, will be realized only once the parity-partner assignment is clarified; \( m_0 \) is tightly constrained through lattice and multi-messenger data; and fluctuation effects are incorporated systematically. Continued cross-disciplinary work—combining effective theory, lattice QCD, heavy-ion phenomenology and neutron-star astrophysics—will determine whether parity doubling represents merely an elegant symmetry construction or a fundamental aspect of strongly interacting matter.

\noindent
A recurring limitation across existing studies concerns the extent to which phenomenological success depends on parameter tuning rather than on unavoidable consequences of parity doubling itself. For example, realistic saturation properties may be reproduced for a wide range of $m_0$ values, yet the corresponding predictions for chiral restoration density, neutron-star radii and cooling behaviour can differ substantially. This raises the question of whether parity doubling alone suffices to constrain dense-matter dynamics, or whether additional dynamical input—such as correlations beyond mean field or explicit quark degrees of freedom—is required for quantitative reliability.

\section{Conclusions}
\label{sec:conclusions}

\noindent
The parity-doublet approach to nuclear and dense matter has evolved from an elegant symmetry-based construction into a broadly employed effective framework with clear phenomenological relevance. Its primary achievement is not the reproduction of nuclear saturation itself—already accomplished within conventional RMF models—but rather the reinterpretation of baryon mass generation in terms of chirally invariant and symmetry-breaking components. Through this conceptual reorganisation, chiral restoration in dense matter can be discussed without requiring baryons to become massless, thereby providing a physically transparent bridge between hadronic phenomenology and QCD expectations.

\noindent
Within modern parity-doublet mean-field formulations, nuclear saturation and in-medium chiral dynamics are obtained consistently from the same underlying Lagrangian. The chiral-invariant mass, $m_0$, once treated as a secondary parameter, has assumed a central phenomenological role: it governs the scalar response of the nucleon and influences the density at which chiral symmetry restoration sets in. Finite-temperature analyses have indicated that, depending on the magnitude of $m_0$ and the strength of the vector couplings, the chiral transition may proceed as a smooth crossover rather than as a pronounced first-order discontinuity. In this way, these models provide a practical and conceptually consistent bridge between purely hadronic descriptions and quark-level effective theories such as the Nambu–Jona-Lasinio or quark–meson coupling frameworks.

\noindent
At the same time, significant theoretical and phenomenological challenges persist. Quantitative determinations of $m_0$ and of the scalar–vector couplings from empirical constraints continue to involve substantial uncertainties. The incorporation of fluctuations beyond the mean-field approximation, especially in the vicinity of the chiral crossover, has not yet been achieved in a systematic manner. From an astrophysical perspective, extensions to the high-density regime relevant for neutron-star interiors require a controlled matching to quark degrees of freedom, potentially through hybrid or parity-doublet–quark hybrid constructions.

\noindent
Several robust conclusions may nevertheless be drawn from the developments reviewed here. First, a sizeable chirally invariant mass $m_0 \approx 600-800$ MeV appears to be simultaneously favoured by nuclear saturation properties, lattice-QCD indications of parity doubling and neutron-star observations. Second, vector interactions remain essential for obtaining realistic equations of state, indicating that chiral symmetry alone is insufficient to determine bulk matter properties. Third, parity-doublet frameworks tend naturally towards smooth crossover behaviour rather than strong first-order transitions, a tendency that is broadly compatible with current lattice and astrophysical constraints.

\noindent
In the near future, further progress is anticipated from two complementary directions: increasingly precise lattice-QCD constraints on in-medium condensates and improved multi-messenger astrophysical measurements of compact-star masses and radii. These inputs are expected to sharpen empirical bounds on chiral-parity mean-field models. The field therefore appears to be at a promising stage—conceptually established, phenomenologically testable and centrally relevant to elucidating the chiral structure of strongly interacting matter.

\section{Acknowledgement}
\label{sec:ack}

\noindent
The author is indebted to the unprecedented support and invaluable insights received from the members of the Department of Physics at Brainware University; in general; and from Dr. Soumyajit Sarkar and Mr. Soham Chandra in particular.

\bibliographystyle{elsarticle-num}
\bibliography{bibnew_stripped_keepdoi}
\end{document}